\documentclass[twocolumn,amsmath,amssymb,pra]{revtex4}
\usepackage{setspace}
\usepackage{graphicx}
\usepackage{amsmath}
\usepackage{amssymb}
\usepackage{latexsym}

\begin{document}

\title{Rapid ramps across the BEC-BCS crossover: a novel route to measuring the superfluid gap}
\author{R.G. Scott$^1$,  F. Dalfovo$^1$, L.P. Pitaevskii$^{1,2}$, S. Stringari$^1$}
\affiliation{$^1$INO-CNR BEC Center and Dipartimento di Fisica, Universit{\`a} di Trento, Via Sommarive 14, I-38123 Povo, Italy.\\$^{2}$Kapitza institute for physical problems, ul. Kosygina 2, 119334 Moscow, Russia.} 
\date{25/8/12}



\begin{abstract}
We investigate the response of superfluid Fermi gases to rapid changes of the three-dimensional s-wave scattering length $a$ by solving the time-dependent Bogoliubov-de Gennes equations. In general the magnitude of the order parameter $\left|\Delta\right|$ performs oscillations, which are sometimes called the ``Higgs'' mode, with the angular frequency $2 \Delta_{\mbox{gap}}/ \hbar$, where $\Delta_{\mbox{gap}}$ is the gap in the spectrum of fermionic excitations. Firstly, we excite the oscillations with a linear ramp of $1/a$ and study the evolution of $\left|\Delta\right|$. Secondly, we continously drive the system with a sinusoidal modulation of $1/a$. In the first case, the oscillations in $\left|\Delta\right|$ damp according to a power law. In the second case, the continued driving causes revivals in the oscillations. In both cases, the excitation of the oscillations causes a reduction in the time-averaged value of $\left|\Delta\right|$. We propose two experimental protocols, based around the two approaches, to measure the frequency and damping of the oscillations, and hence $\Delta_{\mbox{gap}}$.
\end{abstract}

\maketitle

\section{Introduction}

Nearly forty years ago Volkov and Kogan investigated the response of a Fermi superfluid to a small initial perturbation of the order parameter $\Delta$~\cite{volkov}. They predicted weakly damped oscillations in $\left|\Delta\right|$ with the angular frequency $2 \Delta_{\mbox{gap}}/ \hbar$, where $\Delta_{\mbox{gap}}$ is the gap in the spectrum of fermionic excitations. These oscillations originate from the threshold for the creation of fermionic excitations by pair-breaking following a time-dependent perturbation. The threshold causes a branch-type singularity in the related response functions at the frequency $2\Delta_{\mbox{gap}} /\hbar$, and consequently oscillations in $\left|\Delta\right|$ with the same frequency. Hence a detection of the frequency would give a direct measurement of $\Delta_{\mbox{gap}}$. There is no exponential damping because the singularity lies on the real axis. However, since the singularity is not a pole, but a branch-point, the amplitude of the oscillations decreases with time according to a power law. Volkov and Kogan found that, in the Bardeen-Cooper-Schrieffer (BCS) regime, where the fermionic excitations are created with finite momenta $p$ near the Fermi momentum, the amplitude decays as $t^{-1/2}$~\cite{volkov}. More recently, Gurarie calculated that the amplitude decays as $t^{-3/2}$ in the Bose-Einstein condensate (BEC) regime, where the excitations are created near the point $p=0$~\cite{gurarie}. Persistent oscillations have been found for very large initial perturbations and particular nonequilibrium initial states~\cite{yuz2,barankov,bulgac}. Due to a formal analogy between the gap phenomenon in superfluid Fermi liquids and mass creation by the Higgs mechanism in the theory of elementary particles, these oscillations are also known as the ``Higgs mode''~\cite{huberPRA,huber,pollet}. Despite some investigation into the response of a superfluid Fermi gas to a modulation of the scattering length~\cite{Zwierlein,plata}, the Higgs mode has never been observed in Fermi gases. Analogous phenomena have been observed in superconductors~\cite{soo,allvarma} and for Bosons near the Mott-insulator transition~\cite{bloch}. 

It is the purpose of this paper to provide a stepping-stone between the past theory described above and future experiments. By solving the time-dependent Bogoliubov-de Gennes equations~\cite{eagles,leggett,metrento,metrento2,Challis} we investigate the appearance of the order parameter oscillations in superfluid Fermi gases, following rapid ramps across the BEC-BCS crossover. In the spirit of Volkov and Kogan~\cite{volkov}, we begin with relatively slow ramps of the three-dimensional s-wave scattering length $a$ from $a=a_0$ to $a_1$ from $t=0$ to $t = t_1 \approx \hbar/E_f$. We confirm that the angular frequency of the oscillations is given by $2 \Delta_{\mbox{gap}}/ \hbar$, in which $\Delta_{\mbox{gap}} = \left|\Delta\right|$ is the BCS regime and $\Delta_{\mbox{gap}} = \sqrt{\mu^2 + \Delta^2}$ in the BEC regime, where $\mu$ is the chemical potential. The oscillations damp as predicted in Refs.~\cite{volkov,gurarie}. For abrupt ramps such that $t_1 \ll \hbar/E_f$, we find that the order parameter oscillates around a new value $\Delta_\infty$, which is less than the equilibrium value for $a = a_1$. Note that our abrupt ramps never take the order parameter near zero, as instead studied in Ref.~\cite{bulgac}. The frequency of the oscillations is set by the final value of the gap, which in turn depends on $\Delta_\infty$. The value of $\Delta_\infty$ decreases as we increase $\left|1/a_0 - 1/a_1\right|$, see also Refs.~\cite{yuz,yi}. Furthermore, $\Delta_\infty$ depends only on the instantaneous value of $\left|\Delta\right|$ just before the ramp, irrespective of whether the system was in equilibrium or not. We also study the response of $\left|\Delta\right|$ to a sinusoidal modulation of $1/a$. In this case, $\left|\Delta\right|$ performs oscillations about a mean value which gradually decreases with $t$ to a constant (see also Ref.~\cite{yi}). The oscillations do not damp, but show revivals due to the continued driving of the system.  

Building on these findings, we propose two experimental protocols to detect the frequency and damping of the oscillations in $\left|\Delta\right|$ across the BEC-BCS crossover. The first is based on abrupt ramps of $a$, and the second on sinusoidal modulation of $a$. We confirm that the trapping potential in a real experiment would not significantly spoil the effectiveness of either protocol because the density of the gas does not respond on the timescale of the oscillations in $\left|\Delta\right|$.

The paper is organised as follows. In section~\ref{sec:method} we outline the system and methodology. In sections~\ref{sec:slow} and ~\ref{sec:fast} we investigate the response of a uniform Fermi superfluid to slow and abrupt ramps, such that $t_1 \approx \hbar/E_f$ and $t_1 \ll \hbar/E_f$, respectively. In section~\ref{sec:sin} we investigate the response of a uniform Fermi superfluid to a sinusoidal modulation of $1/a$. In section~\ref{sec:exp} (\ref{sec:expB}) we propose experimental protocol A (B), which is based on abrupt ramps of $a$ (sinusoidal modulation of $1/a$). In section~\ref{sec:conc} we conclude.





\section{System and methodology}
\label{sec:method}

We consider a three-dimensional superfluid Fermi gas with equal populations of the two spin components. We model its dynamics across the BEC-BCS crossover by solving the time-dependent Bogoliubov-de Gennes equations~\cite{eagles,leggett,metrento,metrento2,Challis}. Although this is an approximate theory, it is qualitatively correct and able to capture the main physics of the problem~\cite{SandroReview}. The equations are
\begin{equation}
\left[\begin{array}{ll}
\hat{H} & \Delta \mbox(\textbf{r},t) \\
\Delta^* \mbox(\textbf{r},t) & -\hat{H}
\end{array}\right]
\left[\begin{array}{l}
u_\eta \mbox(\textbf{r},t) \\
v_\eta \mbox(\textbf{r},t)
\end{array}\right] = 
i \hbar \frac{\partial}{\partial t}
\left[\begin{array}{l}
u_\eta \mbox(\textbf{r},t) \\
v_\eta \mbox(\textbf{r},t)
\end{array}\right] , 
\label{eq:tdbdg}
\end{equation}
where $\hat{H} = -\hbar^2 \nabla^2 / 2m + U(x) - \mu$, in which $m$ is the atomic mass, $U$ is the external potential and $\mu$ is the chemical potential. The order parameter is calculated as $\Delta\left(\textbf{r},t\right) = -g \sum_{\eta}u_{\eta}v_{\eta}^{*}$, in which $g$ is given by $1/k_f a = 8\pi E_f/ ( g k_f^3 ) + \sqrt{4E_c / \left(\pi^2 E_f\right)}$~\cite{SandroReview}. Here $E_f = \hbar^2 k_f^2 / 2m$ and $k_f = \left(3\pi^2n\right)^{1/3}$ are the Fermi energy and momentum of an ideal Fermi gas of density $n$, respectively. The cut-off energy $E_c$ is introduced in order to remove the ultraviolet divergences in the Bogoliubov-de Gennes equations with contact potentials. The density of the gas is $n \mbox(\textbf{r},t) = 2 \sum_{\eta} \left|v_{\eta} \mbox(\textbf{r},t)\right|^2$. 

\begin{figure}[tbp]
\includegraphics[width=0.5\columnwidth]{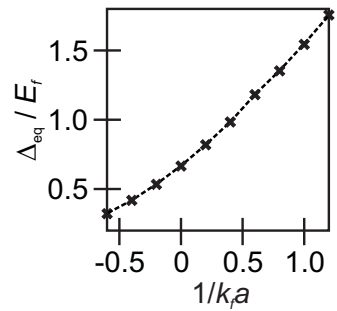} 
\caption{The equilibrium magnitude of the order parameter $\Delta_{\mbox{eq}}\left(1/k_f a\right)$ obtained by solving the stationary Bogoliubov-de Gennes equations for a uniform Fermi gas.}
\label{f0}
\end{figure}

For most of the results in this article, we consider a uniform superfluid in a box with periodic boundary conditions and dimensions $L_x=37.6 k_f^{-1}$ and $L_\bot=12.5 k_f^{-1}$ in the $x$- and other directions respectively. The potential $U=0$. Towards the end of the article, in order to simulate a more realistic system, we also consider the harmonic trapping potential $U=m \omega_x^2 x^2/2$, where $\omega_x$ is the angular trapping frequency in the $x-$direction. Since, throughout the article, the potential $U$ has no $y$ or $z$ dependence, we may write the functions $u_\eta \mbox(\textbf{r},t)$ and $v_\eta \mbox(\textbf{r},t)$ as $u_\eta (x,t) e^{i (k_y y + k_z z)}$ and $v_\eta (x,t) e^{i (k_y y + k_z z)}$ respectively, in which $k_y$ and $k_z$ are quantized according to $k_y = 2\pi \alpha_y / L_\bot$ and $k_z = 2\pi \alpha_z / L_\bot$, where $\alpha_y$ and $\alpha_z$ are integers. We begin our simulations with stationary solutions of Eq.~(\ref{eq:tdbdg})~\cite{metrento,Antezza} for $a = a_0$. Figure~\ref{f0} shows the equilibrium magnitude of the order parameter $\Delta_{\mbox{eq}}\left(1/k_f a\right)$ for a uniform Fermi gas. 

\section{Results and discussion}

\subsection{Slow ramps of the scattering length}
\label{sec:slow}

Our initial goal is to confirm the predictions of Volkov and Kogan~\cite{volkov} and Gurarie~\cite{gurarie} for the period and damping of the oscillations in $\left|\Delta\right|$. In the spirit of Volkov and Kogan, we study the response of the superfluid to a small linear ramp of $1/k_fa$ from $1/k_fa_0$ to $1/k_fa_1$ over the timescale $t_1 \approx \hbar/E_f$. Both the small change in $1/a$ and the relatively large $t_1$ minimise the reduction in $\left|\Delta\right|$, specifically $\Delta_{\mbox{eq}} - \Delta_\infty$, caused by the excitation of the oscillation. In this work, we refer to these ramps as 	``slow'' because they are close to being adiabatic. However, even these ramps are fast enough to be a challenge to realize in experiment.

\begin{figure}[tbp]
\includegraphics[width=1.0\columnwidth]{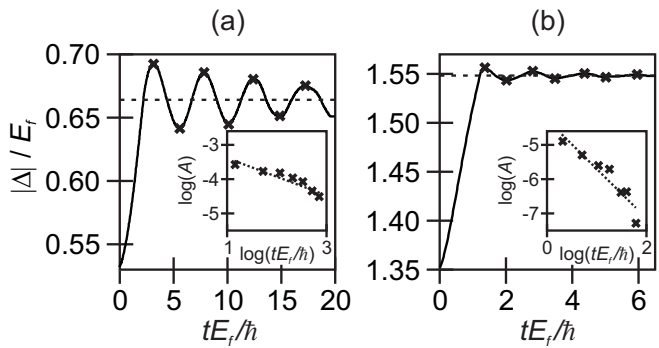} 
\caption{(a) The response of $\left|\Delta(t)\right|$ following a ramp from $1 / k_f a_0 = -0.2$ to $1 / k_f a_1 = 0$ over the time $t_1 = 2.2 \hbar / E_f$. Inset: $\log(A)$ (see text), evaluated at the positions of the crosses in the main figure, against $\log(t)$. The dotted line is a fit with a gradient of $-0.5$. (b) As a, but $1 / k_f a_0 = 0.8$, $1 / k_f a_1 = 1$ and $t_1 = 1.2 \hbar / E_f$. The gradient of the dotted line in the inset is $-1.5$. The two insets are plotted with the same scale to aid comparison.}
\label{f10}
\end{figure}

Figure \ref{f10}(a) shows the evolution of the magnitude of the order parameter $\left|\Delta(t)\right|$ in the uniform superfluid following a ramp from $1 / k_f a_0 = -0.2$ to $1 / k_f a_1 = 0$ (unitarity) over the time $t_1 = 2.2 \hbar / E_f$. After the ramp $\left|\Delta\right|$ oscillates around the average value $\Delta_\infty = 0.664 E_f$ [indicated by the horizontal dashed line in Fig.~\ref{f10}(a)], which is very slightly lower ($0.4\%$) than the equilibrium value at unitarity (see Fig.~\ref{f0}): even this slow ramp of $1/k_f a$ causes some reduction in $\left|\Delta\right|$. The period of the oscillation is $4.7 \hbar/E_f$. This agrees well with the previous prediction of $\pi \hbar / \Delta_{\mbox{gap}}$ by Volkov and Kogan~\cite{volkov}, taking $\Delta_{\mbox{gap}} = \Delta_\infty$. We also quantify the damping of the oscillation by evaluating the function $A(t) = \left|\Delta(t)\right| -  \Delta_\infty $ at the extremes of the oscillation, which are indicated by the crosses in Fig.~\ref{f10}(a). Assuming that $A \propto t^{\gamma}$, we plot $\log(A)$ against $\log(t)$ in the Fig.~\ref{f10}(a) inset and measure the gradient to determine $\gamma$. Our data is consistent with the prediction of Volkov and Kogan~\cite{volkov}, shown by the dotted line, that $\gamma$ and hence the gradient should be $-0.5$.

\begin{figure}[tbp]
\includegraphics[width=1.0\columnwidth]{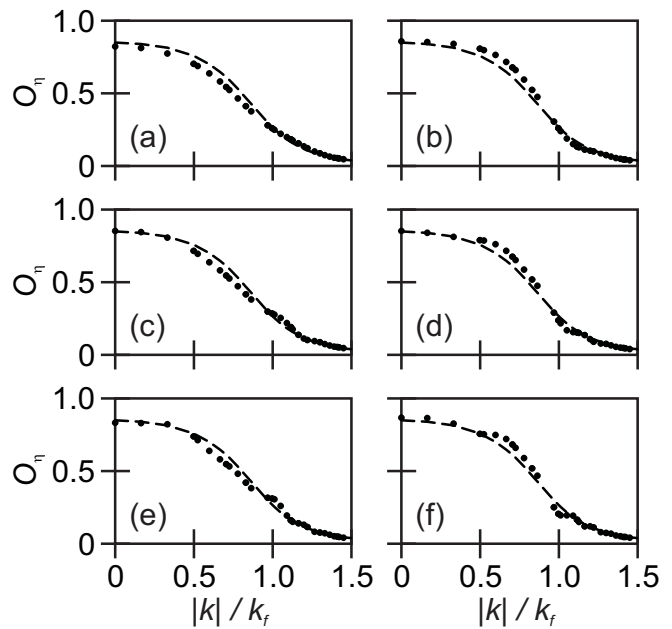} 
\caption{(a) - (f): The filled circles indicate the occupation of the energy levels $O_\eta = \int \left|v_\eta \mbox(\textbf{r},t)\right|^2 d\textbf{r}$ at the points indicated by the crosses in Fig.~\ref{f10}(a), in chronological order from the first maximum. Hence the left-hand (right-hand) column corresponds to maximums (minimums) in $\left|\Delta\right|$. The dashed curve shows the equilibrium distribution function for comparison.}
\label{f1a}
\end{figure}

We may gain a microscopic insight into the oscillations and their damping by studying the occupations of the energy levels $O_\eta = \int \left|v_\eta \mbox(\textbf{r},t)\right|^2 d\textbf{r}$. The occupations of the energy levels at the times indicated by the crosses in Fig.~\ref{f10}(a) are plotted as a function of $\left|k\right|$ with filled circles in Fig.~\ref{f1a}, in chronological order from (a) to (f). Hence the left-hand (right-hand) column corresponds to maximums (minimums) in $\left|\Delta\right|$. (The occupations at the time indicated by the final cross in Fig.~\ref{f10}(a) are not plotted.) For comparison the dashed curve shows the equilibrium distribution function for $1 / k_f a = 0$. We see that the oscillations in $\left|\Delta\right|$ are due to a periodic flatening and steepening of the quasiparticle distribution function. After an increasing number of oscillations, the distribution function becomes less smooth. This is the source of the damping of the oscillations. At large times this effect resembles a thermal excitation of the system, and hence $\Delta_\infty<\Delta_{\mbox{eq}}\left(1/k_f a\right)$. 

To illustrate the corresponding behaviour in the BEC regime, Fig.~\ref{f10}(b) shows the response of $\left|\Delta(t)\right|$ to a ramp from $1 / k_f a_0 = 0.8$ to $1 / k_f a_1 = 1.0$ over the time $t_1 = 1.2 \hbar / E_f$. Following the ramp, $\left|\Delta\right|$ again oscillates around a new value $\Delta_\infty$ [indicated by the horizontal dashed line in Fig.~\ref{f10}(b)], which is very close to but less than the equilibrium value for $1/k_fa = 1$ (see Fig.~\ref{f0}). We measure the period to be $1.6 \hbar/E_f$, which is slightly lower than the predicted $\pi \hbar / \Delta_{\mbox{gap}} = 1.7 \hbar/E_f$, in which $\Delta_{\mbox{gap}}$ is calculated as $\sqrt{\mu^2 + \Delta_\infty^2}$, taking $\mu$ to be the equilibrium value for $1/k_fa = 1$. However, it is difficult to measure the period accurately because the damping is far more rapid than in Fig.~\ref{f10}(a). This is confirmed by the plot of $\log(A)$ against $\log(t)$ in the Fig.~\ref{f10}(b) inset. Our data points now follow the prediction of Gurarie~\cite{gurarie}, shown by the dotted line, that $\gamma$ and hence the gradient is $-1.5$ in the BEC regime.

\subsection{Abrupt ramps of the scattering length}
\label{sec:fast}

The slow ramps of the scattering length in the previous section excite the oscillations in $\left|\Delta\right|$ with a minimal reduction in the time-averaged $\left|\Delta\right|$. However, the amplitude of resulting oscillations in $\left|\Delta\right|$ is quite small, and so it would difficult to base an experimental proposal around this scheme. In this section we discuss the response of the superfluid to an abrupt change in the scattering length. Here ``abrupt'' means short compared to $\hbar/ E_f$. This can excite oscillations in $\left|\Delta\right|$ with a large amplitude, and leads us to an experimental proposal to detect them.
 
\begin{figure}[tbp]
\includegraphics[width=1.0\columnwidth]{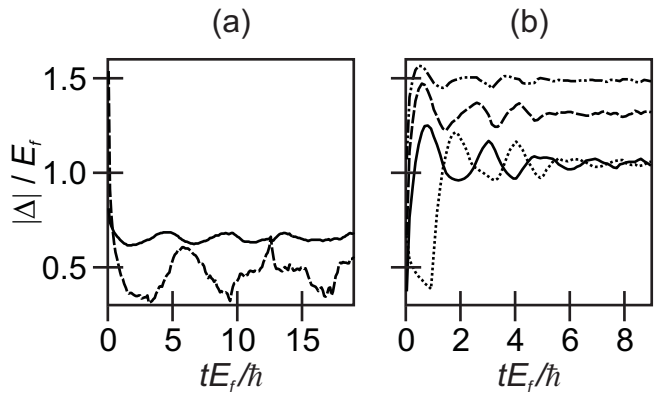} 
\caption{(a) Solid (dashed) curve: the response of $\left|\Delta(t)\right|$ to an abrupt ramp at $t=0$ from $1/k_fa_0=0.2$ ($1.0$) to $1/k_fa_1 = 0$. (b) Solid (dashed/dashed-dotted) curve: as (a), but $1/k_f a_0 = -0.5$ ($0$/$0.5$) and $1/k_f a_1 = 1.0$. Dotted curve: response of $\left|\Delta(t)\right|$ to an abrupt ramp at $t=0$ from $1/k_f a_0=0$ to $1/k_fa_1=-0.5$ and a second at $t=0.90 \hbar/E_f$ to $1/k_fa_2=1.0$.}
\label{f1}
\end{figure}

Firstly, we consider abrupt ramps to $1/k_f a_1 = 0$ (unitarity). Figure~\ref{f1}(a) shows the response of $\left|\Delta(t)\right|$ following an abrupt ramp from $1/k_f a_0 = 0.2$ (solid curve)  and $1.0$ (dashed curve). Both ramps excite oscillations in $\left|\Delta\right|$, but the time-averaged value of the order parameter, $\Delta_\infty$, is smaller for $1/k_f a_0 = 1.0$. In general, we find that $\Delta_\infty$ is always less than the equilibrium value $\Delta_{\mbox{eq}}(1/k_f a_1)$ (see Fig.~\ref{f0}), and that $\Delta_\infty$ is smaller for large ramps. Notice that the period of the oscillations in $\left|\Delta\right|$ is set by $\Delta_\infty$, not by $\Delta_{\mbox{eq}}$. Consequently, the period of the oscillation is longer following the ramp from $1/k_f a_0 = 1.0$. Also note that the damping is more ``noisy'' following an abrupt ramp, so it would be more difficult to measure $\gamma$ by this method.


Now we consider abrupt ramps to $1/k_f a_1 = 1$ (the BEC regime). Figure~\ref{f1}(b) shows the response of $\left|\Delta(t)\right|$ following an abrupt ramp from $1/k_f a_0 = -0.5$ (solid curve), $0$ (dashed curve) and $0.5$ (dash-dotted curve). Again we see that a large ramp leads to a small $\Delta_\infty$, and that as the ramp tends to zero $\Delta_\infty$ approaches $\Delta_{\mbox{eq}}\left(1\right)$. As before, the value of $\Delta_\infty$ sets the period of the oscillations. Since we are in the BEC regime ($1/k_f a=1$) at the end of the simulation, variations in $\Delta_\infty /  \Delta_{\mbox{eq}}\left(1\right)$ cause variations in the condensate fraction, which may be detected through analysis of the bimodal distribution following a free expansion~\cite{jin}. 


The dotted curve shows the response of $\left|\Delta(t)\right|$ to a sequence of two ramps: the first from $1/k_f a_0 = 0$ to $1/k_f a_1 = -0.5$ at $t=0$, and the second to $1/k_f a_2 = 1.0$ at $t=0.90 \hbar/E_f$. We chose this time so that instantaneous value of $\left|\Delta\right|$ just before the second ramp is equal to $\Delta_{\mbox{eq}}\left(-0.5\right)$. We find that $\Delta_\infty$ is identical to that for the single abrupt ramp from the equilibrium superfluid at $1 / k_f a_0 = -0.5$ to $1/k_f a_1 = 1.0$ (solid curve). This illustrates that the superfluid has no memory of its previous dynamics before the abrupt ramp, and that $\Delta_\infty$ is a true measure of the instantaneous value of $\left|\Delta\right|$ just before the ramp. Although we show just one example here, we have confirmed that this is true for other parameters. This is an important requirement for the validity of experimental protocol A, described in section~\ref{sec:exp}.

\subsection{Sinusoidal modulation of the scattering length}
\label{sec:sin}

An alternative method to excite oscillations in $\left|\Delta\right|$ is to modulate $1/k_fa$ at the resonant angular frequency $2 \Delta_{\mbox{gap}}/ \hbar$. This is potentially easier to realize experimentally because $1/k_f a$ would change smoothly. In principle the gap $\Delta_{\mbox{gap}}(t)$ is a function of time, but we always modulate $1/a$ at a constant angular frequency $\omega_m$, according to
\begin{equation}
 1/k_f a\left(t\right) = 1/k_f a_0 + \left(0.04/k_f\right)\sin\left( \omega_m t \right),
\label{eq:sin} 
\end{equation}
In this section we drive at the resonant frequency by setting $\omega_m=2 \Delta_{\mbox{gap}}\left(0\right)$, where $\Delta_{\mbox{gap}}\left(0\right) = \Delta_{\mbox{eq}}\left(1/k_f a_0\right)$ in the BCS regime and $\Delta_{\mbox{gap}}\left(0\right) = \sqrt{\Delta_{\mbox{eq}}^2\left(1/k_f a_0\right) + \mu^2}$ in the BEC regime. The amplitude of the modulation is $0.04/k_f \ll 1/k_f$ so that we probe the response of the superfluid within a narrow range of $1/k_f a$. Moreover, a small modulation of $1/a$ is easier to realize experimentally. The driving continues until $t=t_m=130\hbar/E_f$.  

\begin{figure}[tbp]
\includegraphics[width=0.55\columnwidth]{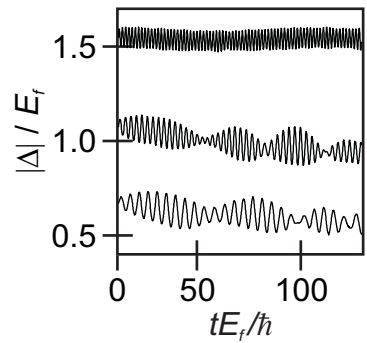} 
\caption{The response of $\left|\Delta(t)\right|$ to a sinusoidal modulation of $1/k_f a$ at the resonant angular frequency $\omega_m = 2 \Delta_{\mbox{gap}}\left(t=0\right)$ [see Eq. (\ref{eq:sin})] with $1/k_f a_0 = 1.0$ (top curve), $1/k_f a_0 = 0.5$ (middle curve) and $1/k_f a_0 = 0.0$ (bottom curve).}
\label{f5}
\end{figure}

Figure~\ref{f5}(a) shows the evolution of $\left|\Delta(t)\right|$ in response to this driving for $1/k_f a_0 = 1.0$, $0.5$ and $0.0$. For $1/k_f a_0 = 1.0$ (top curve), the magnitude of the oscillations increases slightly during the first two oscillations, then plateaus. Further driving causes no further amplification of the oscillations. The time-averaged value of $\left|\Delta\right|$ decreases only very slightly (on the order of $1\%$). This occurs because a small variation of $1/a$ in the BEC regime (large $1/k_f a_0$), where the fermionic atoms are tightly bound in molecules, is not sufficient to break the pairs. The response of $\left|\Delta(t)\right|$ is very different for $1/k_f a_0 = 0.5$ (middle curve) and $1/k_f a_0 = 0.0$ (lower curve). Now we see that the amplitude of the oscillations grows over about five oscillations (see also Ref.~\cite{yi}), as the time-averaged value of $\left|\Delta\right|$ slowly falls. However, on longer timescales we see a beating of the oscillations, and the time-averaged value of $\left|\Delta\right|$ saturates at a constant value. The beating occurs due to an interference between the driving frequency and the frequency of the oscillations in $\left|\Delta\right|$ set by $\Delta_{\mbox{gap}}$.

\subsection{Experimental protocols}
\label{sec:exp}
\subsubsection{Protocol A}
 
\begin{figure}[tbp]
\includegraphics[width=1.0\columnwidth]{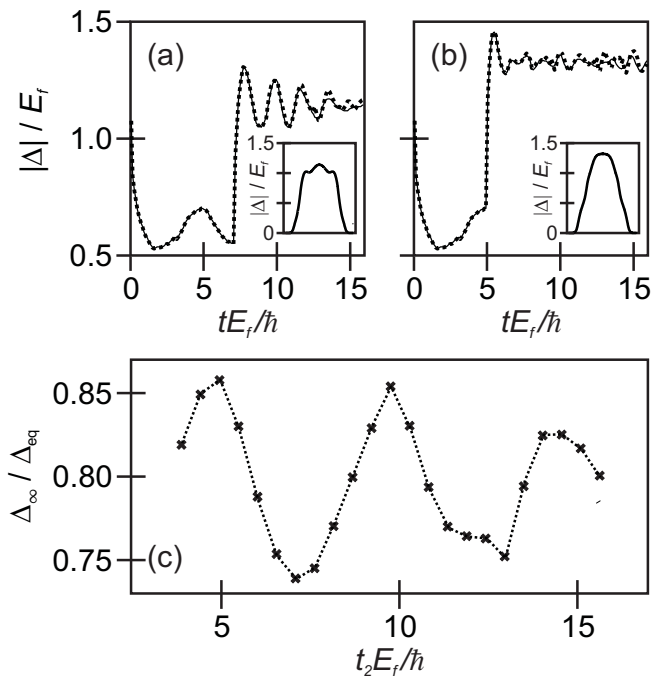} 
\caption{Experimental protocol A to detect the oscillations in $\left|\Delta\right|$. (a) Dotted curve: response of $\left|\Delta(t)\right|$ in a uniform superfluid to an abrupt ramp at $t=0$ from $1/k_fa_0=0.5$ to $1/k_fa_1=0$ and a second at $t=t_2=7.1$ $\hbar/E_f$ to $1/k_fa_2=1.0$. Solid curve: as dotted curve, but for a $^{40}K$ superfluid in a trap with $\omega_x=2 \pi 50$ rad s$^{-1}$ and peak density $1.8 \times 10^{18}$ m$^{-3}$. Inset: profile of $\left|\Delta(x,t=16 \hbar/E_f)\right|$ in the trapped superfluid. The field-of-view is $120 k_f^{-1}$. (b) As (a), but with $t_2=4.9$ $\hbar/E_f$. (c) $\Delta_\infty /  \Delta_{\mbox{eq}}$ in a uniform superfluid as a function of $t_2$.}
\label{f1b}
\end{figure}

Building on the abrupt ramps studied in section~\ref{sec:fast}, we now propose an experimental protocol, which we call protocol A, to measure the frequency and damping of the oscillations in $\left|\Delta\right|$. We perform two abrupt ramps on a superfluid Fermi gas with $1/k_f a_0 = 0.5$. The first at $t=0$ excites the oscillations in $\left|\Delta\right|$ by ramping to $1/k_f a_1 = 0$. The second at $t=t_2$ ramps to $1/k_f a_2 = 1$ (the BEC regime). The dotted curve in Fig.~\ref{f1b}(a) shows the evolution of $\left|\Delta(t)\right|$ in a uniform superfluid for $t_2 = 7.1$ $\hbar/E_f$. In this case, $t_2$ coincides with a trough in $\left|\Delta\right|$, and hence $\Delta_\infty$ is small. The dotted curve in Fig.~\ref{f1b}(b) shows the corresponding simulation for $t_2 = 4.9$ $\hbar/E_f$. In this case, the second ramp coincides with a peak in $\left|\Delta\right|$, and hence $\Delta_\infty$ is large.

In Fig.\ref{f1b}(c) we plot $\Delta_\infty /  \Delta_{\mbox{eq}}\left(1\right)$ as a function of $t_2$. The graph show damped oscillations that directly reflect the behaviour of $\left|\Delta\right|$ at unitarity following the initial ramp. As explained in section~\ref{sec:fast}, in the BEC regime $\Delta_\infty / \Delta_{\mbox{eq}}$ is a measure of the condensate fraction, which may be determined by ballistic expansion~\cite{jin,note3}. Hence this protocol will enable an experimentalist to resolve the frequency of the oscillations in $\left|\Delta\right|$ and their damping across the BEC-BCS crossover. Of course, the experimentalist would actually measure oscillations of $\left|\Delta\right|$ around a mean value which was slightly less than the equilibrium value. Hence the first ramp should be as small as possible, whilst maintaining a large enough variation in $\Delta_\infty$ to be measured in experiment.

The solid curves in Figs.~\ref{f1b}(a) and (b) show the results of the corresponding simulations for a $^{40}K$ superfluid in a harmonic trap, with $\omega_x = 2\pi 50$ rad s$^{-1}$ and peak density $1.8 \times 10^{18}$ m$^{-3}$. Although the oscillations in $\left|\Delta\right|$ are coupled to density oscillations in an inhomogeneous system, $\left|\Delta(t)\right|$ for the trapped superfluid is nearly identical to that for the uniform superfluid because the density cannot change significantly on a timescale of $\hbar/E_f$. The insets show the profile of the order parameter at $t=16 \hbar/E_f$. The profile in the Fig.~\ref{f1b}(a) inset has the smaller peak because $t_2$ coincides with a trough in $\left|\Delta\right|$. Either side of the central peak the profile has some small shoulders, which occur because the frequency of the oscillations in $\left|\Delta\right|$ decreases towards the edge of the cloud where the density and hence $\Delta_{\mbox{gap}}$ is locally smaller. However, this is a minor effect which would only slightly reduce the variation in condensate fraction observed after expansion.

\subsubsection{Protocol B}
\label{sec:expB}

In this section we propose a second experimental protocol, which we call protocol B, based on the periodic modulation of $1/a$ studied in section~\ref{sec:sin}. The advantage of this approach is that $1/k_f a$ would vary smoothly, and hence protocol B would be easier to realize in experiment than protocol A. The disadvantage is that excitation of the superfluid saturates for large driving times $t_m$, as explained in section~\ref{sec:sin}. Hence it could be difficult to strongly excite the superfluid with this method and produce a large experimental signal.

Protocol B is as follows. We excite a superfluid with an initial $1/k_fa = 1/k_f a_0$ using a periodic modulation of $1/a$, as stated in Eq. (\ref{eq:sin}). At time $t=t_m$ we cease driving the superfluid and ramp to $1/k_fa=1$ over the time $t_m<t<t_{m2}$. The timescale $t_{m2}-t_m$ need \textit{not} be short compared to $\hbar/E_f$. The purpose of this ramp is to convert the excited superfluid with $1/k_fa = 1/k_f a_0$ to an excited superfluid in the BEC regime, where the excitation may be quantified from the bimodal distribution. Hence, this final ramp may even be adiabatic. 

\begin{figure}[tbp]
\includegraphics[width=1.0\columnwidth]{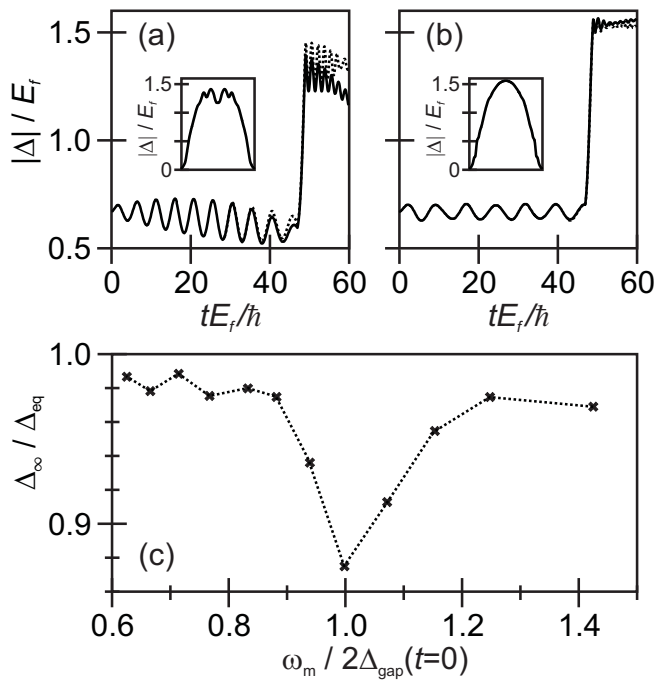} 
\caption{Experimental protocol B to detect the oscillations in $\left|\Delta\right|$. (a) Dotted curve: response of $\left|\Delta(t)\right|$ in a uniform superfluid to a resonant periodic modulation of $a$ [$\omega_m=2 \Delta_{\mbox{gap}}\left(0\right)$, see Eq. (\ref{eq:sin})] followed by a ramp to $1/k_fa=1$ from $t=t_m=47.1$ to $t=t_{m2}=49.1$ $\hbar/E_f$. Solid curve: as dotted curve, but for a $^{40}K$ superfluid in a trap with $\omega_x=2 \pi 50$ rad s$^{-1}$ and peak density $1.8 \times 10^{18}$ m$^{-3}$. Inset: profile of $\left|\Delta(x,t= 60 \hbar/E_f)\right|$ in the trapped superfluid. The field-of-view is $120 k_f^{-1}$. (b) As (a), but for an off-resonant modulation of $a$ [$\omega_m=1.25 \Delta_{\mbox{gap}}\left(0\right)$]. (c) $\Delta_\infty^2 /  \Delta_{\mbox{eq}}^2$ as a function of $\omega_m$.}
\label{f6}
\end{figure}

Figure~\ref{f6} shows an example of protocol B with $1/k_f a_0 = 0$, $t_m = 10 \pi \hbar / \Delta_{\mbox{gap}}(0) = 47.1 \hbar/E_f$ and $t_{m2} - t_m = 2.0 \hbar/E_f$. The dotted curve in Fig.~\ref{f6}(a) shows the evolution of $\left|\Delta(t)\right|$ in a uniform superfluid for resonant driving [$\omega_m=2 \Delta_{\mbox{gap}}\left(0\right)$, see Eq. (\ref{eq:sin})], whilst the dotted curve Fig.~\ref{f6}(b) shows the corresponding result for off-resonant driving ($\omega_m=1.25 \Delta_{\mbox{gap}}\left(0\right)$). Following the ramp into the BEC regime, the value of $\Delta_\infty$ is much smaller for $\omega_m=2 \Delta_{\mbox{gap}}\left(0\right)$, indicating that the resonant driving more strongly excites the superfluid. As explained previously, variations in $\Delta_\infty /  \Delta_{\mbox{eq}}$ in the BEC regime may be detected by measuring the condensate fraction from the bimodal distribution following a free expansion. In Fig.~\ref{f6}(c) we plot $\Delta_\infty / \Delta_{\mbox{eq}}\left(1\right)$ for a uniform superfluid as a function of $\omega_m$. The minimum about $\omega_m=2 \Delta_{\mbox{gap}}\left(0\right)$ indicates the greater excitation of the gas at the resonant frequency of modulation.  

The solid curves in Figs.~\ref{f6}(a) and (b) show the results of the corresponding simulations for a $^{40}K$ superfluid in a trap, with $\omega_x = 2\pi 50$ rad s$^{-1}$ and peak density $1.8 \times 10^{18}$ m$^{-3}$. The solid and dotted curves in Fig.~\ref{f6}(a) diverge slightly for $t>40 \hbar/E_f$ because the density profile of the trapped superfluid begins to respond to the excitation of $\left|\Delta\right|$. This is illustrated by the Figs.~\ref{f6}(a) and (b) insets, which show the profiles of $\left|\Delta(x,t=60 \hbar/E_f)\right|$. The profile in the Fig.~\ref{f6}(a) inset has a central dip. This occurs because the gas is most strongly excited in the center where the frequency of the modulation is resonant. Towards the edges of the cloud the density is lower, and hence the local $\Delta_{\mbox{gap}}$ is lower, meaning that the frequency of the modulation is larger than the local resonant frequency.    


\section{Conclusions}
\label{sec:conc}


Using Bogoliubov-de Gennes theory we have explored how the order parameter $\Delta$ in a superfluid Fermi gas responds to rapid ramps across the BEC-BCS crossover. We have studied both linear ramps and sinusoidal modulations of $1/a$, where $a$ is the three-dimensional s-wave scattering length. In general $\left|\Delta\right|$ performs oscillations with the angular frequency $2 \Delta_{\mbox{gap}}/ \hbar$ about a mean value $\Delta_\infty$, which is less than the equilibrium value of $\left|\Delta\right|$. These oscillations are sometimes called the ``Higgs mode'' due to a formal analogy between the gap phenomenon in superfluid Fermi liquids and mass creation by the Higgs mechanism in the theory of elementary particles~\cite{huberPRA,huber,pollet}. The oscillations damp according to a power law following a linear ramp of $1/a$. Continuous modulation of $1/a$ causes revivals of the oscillations.

The main purpose of this work is to propose experimental protocols to detect the oscillations in $\left|\Delta\right|$ in Fermi superfluids, and hence measure $\Delta_{\mbox{gap}}$. We have proposed two protocols, refered to as protocol A and protocol B, which are based on abrupt ramps of $a$ and modulation of $1/a$, respectively. Protocol A requires the experimentalist to change the scattering length on a timescale faster than $\hbar/E_f$. This is challenging, but may be achieved by using coils with low inductance, or by minimising $E_f$ with dilute clouds or heavy atoms such a Yb~\cite{yb}. Protocol B is potentially easier to realize in experiment, because $1/k_fa$ would vary smoothly. Protocol A has the advantage that it could quantify the damping of the oscillations. However, it has the disadvantage that it would measure the $\Delta_{\mbox{gap}}$ following an abrupt ramp, and consequently the measured $\Delta_{\mbox{gap}}$ may be significantly lower than the equilibrium value. In contrast, Protocol B excites the equilibrium superfluid with a small-amplitude modulation of $1/k_fa$, and hence may accurately measure the $\Delta_{\mbox{gap}}$ of the equilibrium gas.





\bibliography{biblio}

\begin{thebibliography}{25}
\bibitem[{not()}]{note3}
\bibinfo{note}{A short hold time, of the order of the trap period, may be
  required to allow the condensed and uncondensed molecules to thermalise
  within the trap before the free expansion.}
\end{thebibliography}

\end{document}